\documentclass
{iopart}
\usepackage{bbm,iopams,ulem}
\usepackage{graphicx,psfrag}
\usepackage{hyperref}

\newcommand{\ket}[1]{|#1\rangle}
\newcommand{\bra}[1]{\langle#1|}
\newcommand{\mv}[1]{\left\langle#1\right\rangle_\mathrm{av}}
\newcommand{\psop}[2]{\left(\left.#1\right|#2\right)}
\newcommand{\psopdot}[1]{\left(\left.#1\right| \ \cdot\ \right)}
\newcommand{\cg}[2]{\bra{#1}#2\rangle}
\newcommand{\eq}[1]{\begin{equation} #1 \end{equation}}
\newcommand{\eqlab}[2]{\begin{equation} #1
        \label{#2.eq}\end{equation}}
\newcommand{\refeq}[1]{(\ref{#1.eq})}
\newcommand{\calL}{\mathcal{L}}
\newcommand{\calT}{\mathcal{T}}
\newcommand{\Jg}{J_\mathrm{g}}
\newcommand{\Je}{J_\mathrm{e}}
\newcommand{\sfG}{\mathsf{G}}
\newcommand{\bmr}{\bi{r}}
\newcommand{\bQ}{\bi{Q}}

\newcommand{\abcd}{_{\alpha\beta,\gamma\delta}}
\newcommand{\sixj}[6]{\left\{\begin{array}{ccc}
                #1      & #2    & #3\\
                #4      & #5    & #6
                \end{array}\right\}}
\newlength{\argwidth}
\newlength{\vertexheight}
\newcommand{\vertex}[4]{
    \settowidth{\argwidth}{$#1$}
    \hspace{\argwidth}
    \raisebox{\vertexheight}
                {%
        \begin{picture}(25,25)(0,0)
            \put(0,0){\line(1,0){25}}
            \put(0,25){\line(1,0){25}}
            \put(8.5,0){\line(0,1){25}}
            \put(16.5,0){\line(0,1){25}}
        \put(0,25){\hspace{-\argwidth}\raisebox{-0.5ex}{$#1$}}
        \put(25,25){\raisebox{-0.5ex}{$#2$}}
        \put(25,0){\raisebox{-0.5ex}{$#3$}}
        \settowidth{\argwidth}{$#4$}
        \put(0,0){\hspace{-\argwidth}\raisebox{-0.5ex}{$#4$}}
    \end{picture}}
    \settowidth{\argwidth}{$#3$}
    \hspace{\argwidth}
}
\newcommand{\twistedvertex}[4]{
    \settowidth{\argwidth}{$#1$}
    \hspace{\argwidth}
    \raisebox{\vertexheight}{%
        \begin{picture}(25,25)(0,0)
            \put(0,0){\line(1,0){25}}
            \put(0,25){\line(1,0){25}}
            \qbezier(8.5,0)(8.5,6.25)(12.5,12.5)
        \qbezier(12.5,12.5)(16.5,18.75)(16.5,25)
            \qbezier(8.5,25)(8.5,18.75)(11.86,13.5)
        \qbezier(13.14,11.5)(16.5,6.25)(16.5,0)
        \put(0,25){\hspace{-\argwidth}\raisebox{-0.5ex}{$#1$}}
        \put(25,25){\raisebox{-0.5ex}{$#2$}}
        \put(25,0){\raisebox{-0.5ex}{$#3$}}
        \settowidth{\argwidth}{$#4$}
        \put(0,0){\hspace{-\argwidth}\raisebox{-0.5ex}{$#4$}}
    \end{picture}}
    \settowidth{\argwidth}{$#3$}
    \hspace{\argwidth}
}

\begin{document}

\title{Mesoscopic scattering of spin $s$ particles}

\author{C A M\"uller$^1$, C Miniatura$^2$, E Akkermans$^3$ and G Montambaux$^4$}
\address{$^1$ Physikalisches Institut, Universit\"at Bayreuth, 95440
Bayreuth, Germany}
\address{$^2$ Institut Non Lin\'eaire de Nice, UMR 6618 du CNRS, 1361 route des Lucioles, F-06560 Valbonne, France}
\address{$^3$ Department of Physics, Technion, 32000 Haifa, Israel}
\address{$^4$ Laboratoire de Physique des Solides, Universit\'e
Paris-Sud, 91405 Orsay Cedex, France}

\begin{abstract}
Quantum effects in weakly disordered systems are governed
by the properties of the elementary interaction between propagating particles and
impurities. Long range mesoscopic effects due to multiple scattering
are derived by iterating the single scattering vertex, which has to be appropriately
diagonalized. In the present contribution, 
we present a systematic and detailed diagonalisation of the diffuson
and cooperon vertices responsible for weak localization effects. 
We obtain general expressions
for eigenvalues and projectors onto eigenmodes, for any spin and
arbitrary elementary interaction with impurities. This description
provides a common frame for a unified theory of mesoscopic
spin physics for electrons, photons, and other quantum particles. We
treat in detail the case of spin-flip scattering of electrons by freely
orientable magnetic impurities and briefly review the case of photon
scattering from degenerate dipole transitions in cold atomic gases. 
\end{abstract}


\section{Introduction}

The physics of multiple scattering is governed by 
the iteration of elementary scattering events. 
The description of mesoscopic effects, due to phase coherent multiple 
scattering of waves, therefore requires the 
elementary interaction to be in a form which is suitable for the 
iteration. 
The scattered particles in question may be electrons, photons, neutrons, or cold
atoms;  the scatterers may be point-like impurities, spin-flip
impurities interacting with the electron spin via an exchange
interaction,  spin-orbit impurities, atoms interacting with the
photons via the dipolar interaction, or classical dielectric
light scatterers \cite{Gilles&Eric}, to name a few.  

On a classical level, multiple scattering is described by a
Boltzmann-type transport equation, which in a microscopic description  
is generated by considering  pairs of 
complex conjugate amplitudes co-propagating along the same
scattering path. In a diagrammatic representation, these amplitudes
are depicted by the so-called ladder diagrams. The sum of these ladder
diagrams  constitutes the ``diffuson'' which, in the long-distance limit, describes a diffusion process.
Weak localization corrections to classical diffusive transport 
are described by the ``cooperon'', the sum of so-called maximally crossed diagrams made of
amplitudes that are counter-propagating along the same scattering path. 
If the wave scatters off structureless point scatterers, the maximally
crossed diagrams can be disentangled by returning one amplitude and
thus transform into a sum of ladder diagrams. Then, reciprocity
\cite{BvT97} assures that the quantum correction due to this
interference is maximal.  

In the presence of scatterers with internal degrees of freedom, 
multiple scattering contains richer physics, since
these   internal degrees of freedom couple to the degrees of freedom of the
propagating wave. 
The two subsytems, propagating wave and
impurities, become entangled, and discarding all which-path information by
tracing out the unobservable impurity degrees of freedom leads 
to an effective dephasing of coherent effects for the observed wave.
In mesoscopic electronic samples, for example, the spin of a propagating
electron couples to the spin of magnetic impurities, and in
light-scattering atomic clouds the
polarization of propagating photons interacts with the internal
atomic angular momentum. In these
cases, the elementary scattering vertex in the diffuson series is
a tensor with 4 spin indices, connecting 
two incident spin states to two scattered spin states. 
A succesful derivation of multiple scattering properties then requires
that this elementary vertex be iterated.  This problem has been
studied and solved in several specific cases. For instance, in the case of 
spin-orbit coupling and scattering by magnetic impurities,
Hikami, Larkin and Nagaoka \cite{Hikami80} showed that the
cooperon can be diagonalized in the singlet and triplet subspaces. 
Similar methods were employed for calculating conductance fluctuations
\cite{ucfso}. 
In the context of light scattering by thermal atomic gases, the
iteration structure of the photonic diffuson was completely determined
by Barrat, Omont and others  
\cite{barrat, omont, perel}. For the study of phase-coherent effects in cold
atomic gases, the cooperon for the atom-photon problem was calculated
exactly by two of us \cite{Mueller02}. Its diagonal properties were
used to describe coherent backscattering by cold
atomic gases and weak localization phase coherence times \cite{amm}. 
In all these cases,  the iteration of the elementary vertex properties was done
by hand, finding the appropriate diagonal tensors and associated
eigenvalues 
rather heuristically.

In the present contribution, we provide a thorough understanding of
the diffuson and cooperon vertex
diagonalization, of the different projectors and eigenvalues
involved, of their spin and coupling dependence as well as of
their precise relationships.  In the
next section we start by recalling the example of the spin-flip
scattering of spin $\frac{1}{2}$ particles.
In section \ref{projectors.sec}, we present a
general diagonalization scheme for arbitrary ladder and crossed
vertices. We derive in detail the isotropic projectors onto invariant subspaces for
scalar vertices. 
Once the algebraic structure has thus been laid, we calculate in section
\ref{eigenvalues.sec} the correponding eigenvalues from the
microscopic scattering potential. 
Finally, we conclude this paper by indicating some possible extensions of the
work. \ref{photon.sec} contains a brief review of photon scattering
properties in the light of the present work. 

\section{A heuristic diagonalization of the electronic spin-flip
vertex}

\subsection{Diffuson and cooperon} 
\label{probadiff.sec} 

Let us first consider multiple scattering of a quantum particle by
randomly distributed 
impurities without internal
structure \cite{Gilles&Eric}. The disorder-averaged probability of a wavepacket
emanating at point $\bmr$ and time $t=0$ with density matrix
$\rho_0(\bmr)$  to be detected at point
$\bmr'$ and time $t>0$ is given by 
$P(\bmr,\bmr',t) = \mv{ \bra{\bmr'}U(t)\rho_0(\bmr)U^\dagger(t)
\ket{\bmr'}}$ where $\mv{\dots}$ signifies a trace over the impurity
configurations. 
For a quasimonochromatic wave packet of central energy $\varepsilon$
and long evolution time, the Fourier transform  of the detection
probability is  
$P(\bmr,\bmr',\omega) = [2\pi\rho(\varepsilon)]^{-1} \mv{G^\mathrm{R}(\bmr,\bi
r',\varepsilon)G^\mathrm{A}(\bmr,\bmr',\varepsilon-\omega)}
$ 
in terms of the retarded and
advanced Green functions $G^\mathrm{R,A}(E)$ and the average density of states
$\rho(\varepsilon)$. 

This average probability satisfies an integral equation of the
Bethe-Salpeter type generating a multiple scattering sequence. In
weakly disordered samples, the interference of amplitudes propagating
along different scattering paths will be washed out by the disorder
average. Therefore, the dominant contribution will come from
co-propagating amplitudes along identical scattering paths, thus
discarding interference effects and recovering a classical propagation
picture. This propagation is described by the so-called diffuson, the
propagation kernel of the multiple scattering sequences defined in
operator form by  
$D = L + L  \sfG D$
where $L$ describes the elementary scattering by a single impurity, 
and the four-point operator $\sfG = \mv{G^\mathrm{A}}\mv{G^\mathrm{R}}$ is the intensity 
propagator (of Boltzmann-Drude type with factorized averages) between scattering
events. 
In diagrammatic representations, this series has a ladder structure
and can formally be summed as a geometric series,  $D =
L/(1-\sfG L)$. Going to the diffusion approximation (Kubo-limit
of large distances and long times) permits to derive the effective diffusion
constant of this classical diffusion process as function of the
microscopic parameters, in a spirit similiar to the kinetic equation in the
Boltzmann-Lorentz model of classical particles colliding with fixed
impurities. In the case of electrons, the Einstein relation between
the diffusion constant and the conductivity then allows to recover the
classical Drude conductivity.

\begin{figure}
\begin{center}
\psfrag{(a)}{\textbf{(a)}}
\psfrag{(b)}{\textbf{(b)}}
\psfrag{(c)}{\textbf{(c)}}
\includegraphics[width=0.75\textwidth]{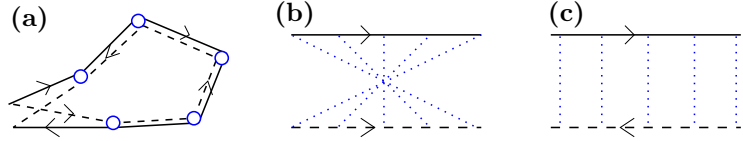}
\caption{Three equivalent representations of the cooperon interference
at five scatterers: 
\textbf{(a)} Real-space representation of counter-propagating
amplitudes;  
\textbf{(b)} momentum-space maximally crossed diagram with retarded
propagator (upper full line) and advanced propagator (lower dashed line)
connected by impurity scattering events (dotted lines);
\textbf{(c)} the same diagram with returned advanced propagator line
exhibiting the ladder structure.} 
\label{cooperon.fig}
\end{center}
\end{figure}

Quantum corrections generated by the interference of amplitudes
propagating along different scattering paths can be incorporated by
considering more general scattering processes. A particular
interference that survives the disorder average is the weak
localization correction of the classical diffusion constant due to
amplitudes counter-propagating along identical scattering paths
\cite{bergmann84}, 
depicted in figure \ref{cooperon.fig}(a).  
In a diagrammatic represention, this interference is given by
maximally crossed diagrams [figure \ref{cooperon.fig}(b)] that can be
unfolded to a ladder structure [figure \ref{cooperon.fig}(c)]
and summed to $C=X/(1-\sfG X)$ where $X$ is the single scattering
vertex for the returned advanced amplitude line. 
For simple impurities without internal structure, one has $X=L$. 
Weak localization then enhances the classical return
probability of a particle by a factor of 2 and reduces the diffusion
constant. Experimentally, this effect can be measured for instance in
the electronic negative magnetoresistance where an external magnetic
field suppresses the weak localization corrections and thus leads to a
larger conductance. In optics, the interferential enhancement of
backscattered intensity is called coherent backscattering and has been observed in a large variety of
samples. The case of impurities with internal structure acting on the
spin degrees of freedom of the propagating particle is more involved
as should become apparent in the following example of electronic 
spin-flip scattering.

\subsection{Definition of the spin-flip vertex}

We consider 
now 
a particle of spin $\bi S$ propagating in
a disordered sample with 
spin $\bi J$ magnetic impurities.
The
particle spin states are written as eigenstates $\ket{s\alpha}$ of
$\bi S^2$ and $S^z$. For electrons, $s=\frac{1}{2}$ and $\alpha=\pm \frac{1}{2}$.
The spin operator then is $\bi S = \bsigma/2$  (in natural units
$\hbar=1$) where the components of 
$\bsigma = (\sigma^x,\sigma^y,\sigma^z)$ are the usual Pauli
matrices. The interaction with a given magnetic
impurity is described by the hermitian operator $V_{m} = g
\, \bi J \cdot \bi S $ with coupling strength $g$.
The Born scattering amplitude for the spin-flip process
$\ket{s\alpha}\mapsto \ket{s\gamma}$ is
\eqlab{
    \bra{s\gamma} V_{m} \ket{s\alpha}
    =  g \, \bi J\cdot \bra{s\gamma}\bi S\ket{s\alpha}
    = g \, \bi J\cdot \bi S_{\gamma\alpha} \
.} {Vm}
The effective scattering intensity of this process is described by
the four-point vertex  (the
overline indicates complex conjugation)
\eqlab{
\fl
\calL\abcd
= \vertex{\alpha}{\gamma}{\delta}{\beta} = \mv{
\bra{s\gamma}V_m\ket{s\alpha} \overline{\bra{s\delta}V_m\ket{s\beta}}}
    = \frac{|g|^2\,J(J+1)}{3}\,\bi S_{\gamma\alpha}\cdot\bi
    S_{\beta\delta}
}{defV0.spin-flip}
where $\mv{\dots}$ denotes a trace over the impurity configurations,
here an isotropic
average  $\mv{J_iJ_j} = \delta_{ij}J(J+1)/3$
over all possible orientations of the freely orientable
magnetic impurity.
This average is the fundamental reason for the  non deterministic dephasing of
the  multiple
 scattering process.
We choose to normalise the spin-flip scattering strength $g$ such
that the intensity vertex is written as
\eqlab{
  \calL\abcd = \frac{\bi S_{\gamma\alpha}\cdot\bi S_{\beta\delta}}{s(s+1)}.
}{defV.spin-flip}
This normalisation choice endows the vertex with the
convenient trace-preserving property $\calL_{\alpha\alpha,\gamma
\delta} = \delta_{\gamma\delta}$; here as in the following,
summation over repeated spin indices is understood.

Weak localization corrections to
transport are embodied in the so-called cooperon and are generated by maximally crossed
diagrams.
These diagrams can be
``unfolded'' to a ladder structure such that
the intensity vertex \refeq{defV.spin-flip} is replaced by the crossed
vertex
\eqlab{
X_{\alpha\beta,\gamma\delta} = \calL_{\alpha \delta,\gamma\beta}
=\twistedvertex{\alpha}{\gamma}{\delta}{\beta}
= \bra{s\gamma\, s\delta}X\ket{s\alpha\, s\beta} = \frac{\bi
S_{\gamma\alpha}\cdot\bi S_{\delta\beta}}{s(s+1)} \ .
}{defX.spin-flip}
For reasons that will become clear later, the crossed
vertex $X$ is denoted by a roman letter, whereas the vertex
$\calL$ is held in curly script.

\subsection{Elementary diagonalization}
\label{heur.diag.sec}

The natural coupling scheme for the vertices  \refeq{defV.spin-flip} and \refeq{defX.spin-flip}
is the ``vertical'' combination
$(\alpha\gamma)\leftrightarrow(\beta\delta)$ between the
elementary scattering amplitudes. However,
in multiple scattering diagrams, the above intensity vertices
have to be chained ``horizontally'' in the direction
$(\alpha\beta)\leftrightarrow (\gamma\delta)$ according to the
following product rule of four-point vertices:
\eqlab{
    (\mathsf{G}\calL)\abcd
        = \calL_{\alpha\beta,\mu\nu} \mathsf{G}_{\mu\nu,\gamma\delta}
         \
}{defproduit}
In the product definition,
the order of operators is inverted since operators are
conventionally applied to their arguments from the left, but their
vertex symbols are usually added to a diagram on the right.
The rank-four tensor $\mathsf{G}\abcd =
\mv{G^\mathrm{R}_{\alpha\gamma}}\mv{G^\mathrm{A}_{\beta\delta}}$ describes the
average propagation between scattering events. 
By virtue of rotational invariance, the average propagators
$\mv{G_{\alpha\gamma}}=\mv{G}\delta_{\alpha\gamma}$ are proportional to the 
identity in spin space such that  
$\mathsf{G}\abcd =
\mv{G^\mathrm{R}}\mv{G^\mathrm{A}}\delta_{\alpha\gamma}\delta_{\beta\delta}$
is proportional to the ``horizontal'' identity.
(Note that for photons, however, transversality
implies that $\mathsf{G}\abcd$ is not the identity which leads to the more
complicated scenario described in appendix A, featuring nonetheless the
general properties discussed in the present section.)
But in order to calculate the summed diffuson 
$\mathcal{D} =\mathcal{L}/(\mathbbm{1}-\mathsf{G}\mathcal{L})$ and the
cooperon $C = X/(\mathbbm{1}-\mathsf{G}X)$, the vertices $\calL\abcd$ and $X\abcd$
have to be diagonalized with respect to the horizontal direction
$(\alpha\beta)\leftrightarrow (\gamma\delta)$.
For electrons with $s=\frac{1}{2}$, this
amounts to diagonalising $4\times 4$ matrices
\cite{Gilles&Eric}.
It turns out that the diffuson and cooperon
vertices can be cast in the form
\begin{eqnarray}
\calL\abcd & = & 
 \lambda_0 \,\calT^{(0)}\abcd  + \lambda_1 \,\calT^{(1)}\abcd,
\label{decompL.spin-flip.eq}\\
X\abcd & =& 
\chi_0 \, T^{(0)}\abcd + \chi_1 \, T^{(1)}\abcd.
\label{decompX.spin-flip.eq}
\end{eqnarray}
Here, the diffuson vertex tensors
\begin{eqnarray}
    \calT^{(0)}\abcd & = &\frac{1}{2}
                   \, \delta_{\beta\alpha}\delta_{\gamma\delta},
\label{calT0.electron.eq}\\
    \calT^{(1)}\abcd & = &\frac{1}{2}
    \, \bsigma_{\beta\alpha}\cdot\bsigma_{\gamma\delta}
    =  \delta_{\gamma\alpha}\delta_{\beta\delta} -
          \frac{1}{2}\, \delta_{\beta\alpha}\delta_{\gamma\delta}
\label{calT1.electron.eq}
\end{eqnarray}
are orthogonal projectors with respect to the horizontal product rule
\refeq{defproduit}:
\eq{
\calT^{(K)}_{\alpha\beta,\mu\nu}\calT^{(K')}_{\mu\nu,\gamma\delta}
   =\delta_{KK'}\calT^{(K)}\abcd.
}
Likewise, the cooperon vertex tensors
\begin{eqnarray}
    T^{(0)}\abcd   & = &  \frac{1}{2} \,
   (\delta_{\gamma\alpha}\delta_{\delta\beta} -
     \delta_{\delta\alpha}\delta_{\gamma\beta}),
\label{T0.electron.eq} \\
    T^{(1)}\abcd  & = & \frac{1}{2} \, (\delta_{\gamma\alpha}\delta_{\delta\beta} +
\delta_{\delta\alpha}\delta_{\gamma\beta}) 
\label{T1.electron.eq}
\end{eqnarray}
are orthogonal projectors such that
$T^{(K)}_{\alpha\beta,\mu\nu} T^{(K')}_{\mu\nu,\gamma\delta}
   =\delta_{KK'} T^{(K)}\abcd$.
Both sets of projectors sum up to the identity 
$\delta_{\gamma\alpha}\delta_{\beta\delta}$ for the
horizontal product rule \refeq{defproduit}.  
Obviously, the diffuson projectors \refeq{calT0.electron} and
\refeq{calT1.electron}  are different from the cooperon projectors 
\refeq{T0.electron} and  \refeq{T1.electron}. This is in sharp contrast to the case of
photon scattering ($s=1$) by atoms with degenerate  dipole transitions,
where the \textit{same} set of orthogonal projectors can be used for both
vertex types \cite{Mueller02}
(see \ref{photon.sec} for details).

In the diagonal decomposition \refeq{decompL.spin-flip}, the
eigenvalues of the diffuson vertex are found to be
$\lambda_0=1$ (non-degenerate) and $\lambda_1= -\frac{1}{3}$
(3-fold degenerate).
The eigenvalues of the normalised crossed vertex  
are $\chi_0=-1$ (non-degenerate) and $\chi_1=
\frac{1}{3}$ (3-fold degenerate).
It has been noticed that the cooperon
spin vertex eigenvalues $\chi_K$ correspond to the singlet channel
$K=0$ and the triplet channel $K=1$, respectively, which accounts
for the degeneracies
\cite{Hikami80}.
 Remarkably, the eigenvalues
$\lambda_K$ and $\chi_K$ are equal in magnitude but opposite in sign, which is not
properly explained on this level of heuristic diagonalization.

Prompted by these observations, we wish to answer the following questions:
\begin{enumerate}
\item
Given diffuson and cooperon vertices for arbitrary spin $s$,
which are the orthogonal projectors that assure a least redundant diagonalization?

\item
How do the diffuson and cooperon eigenvalues depend on the
microscopic spin scattering mechanism?
\end{enumerate}

\subsection{First answers}

\label{essence.sec}

\subsubsection{General idea}

In essence, the intensity vertices 
map two incident spins $s$
onto two final spins. Furthermore, they are scalar objects since
they are  obtained by an isotropic average over microscopic degrees of freedom.
The invariance
under rotations is then responsible for the eigenvalue degeneracies.
The key idea is to decompose the argument and image spaces into irreducible subspaces with respect to the rotation group.
The relevant subspaces are labelled by
the effective recoupled spin $K=0,\dots, 2s$.
A generic scalar vertex $A$ can only connect irreducible
subspaces with equal $K$, and its eigenvalues are degenerate in each subspace.
In the appropriate recoupled basis, a scalar vertex takes the
diagonal form
\eqlab{ A = \left( \begin{array}{cccc}
    a_0\mathbbm{1}_0 &  0 & \dots  & 0 \\
     0  & a_1\mathbbm{1}_1 & \dots & 0  \\
     \vdots  & \vdots & \ddots & \vdots \\
     0  & 0 & \dots & a_{2s}\mathbbm{1}_{2s}
    \end{array} \right) 
= \sum_{K=0}^{2s} a_K T^{(K)} .
}{adiagonal} 
The projectors $T^{(K)}$ are simply
projectors onto the irreducible subspaces. 
Therefore, rotational symmetry alone
dictates that there are at most ($2s+1$) different eigenvalues,
each $(2K+1)$-fold degenerate. This decomposition is optimal if
the scalarity is the only information available and holds for
arbitrary spin.

\subsubsection{Recoupling schemes} 

Clearly, the natural coupling between spin indices in the ladder
and crossed vertices \refeq{defV0.spin-flip} and \refeq{defX.spin-flip} is the
``vertical'' coupling scheme
$(\alpha\gamma)\leftrightarrow(\beta\delta)$ that is 
 inherited from the scattering amplitude \refeq{Vm}. Unfortunately,
this coupling is not suited for an iteration with the product
\refeq{defproduit}. For the diagonalisation, we therefore have to
recouple the spin indices: into the ``horizontal'' coupling scheme  
$(\alpha\beta)\leftrightarrow(\gamma\delta)$ for the ladder vertex,
and the  ``diagonal'' coupling scheme  
$(\alpha\delta)\leftrightarrow(\beta\delta)$ for the crossed vertex. 
Consequently, all vertex eigenvalues we derive will feature
$6j$-symbols that describe the recoupling
of 4 spins in  angular momentum theory.

By exchanging the indices $ \delta \leftrightarrow \beta$ in the
crossed vertex, we
are actually able to recover formally the ladder structure, but there
is a price to be paid. 
Taking seriously the disposition of kets and bras, we see that the two
vertices have different rotational structure: the crossed vertex
defines a linear mapping between product states
$$X : \qquad
\ket{s\alpha}\ket{s\beta}\mapsto \ket{s\gamma}\ket{s\delta} \ ,$$
whereas the diffuson vertex is a mapping not between states, but between
operators:
$${\cal L} : \qquad \ket{s\alpha}\bra{s\beta}\mapsto
\ket{s\gamma}\bra{s\delta} \ .$$
This difference, due to the exchange $\bra{s\beta}
\leftrightarrow \ket{s \delta}$ 
and subsequent relabeling $\beta\leftrightarrow\delta$ 
in the unfolding procedure of the
crossed to the ladder series,  
leads eventually to two distinct sets of projectors.
We will therefore diagonalise the diffuson vertex as a
{\it superoperator},
but treat the cooperon vertex as an ordinary operator.

\subsubsection{Projectors}

In section \ref{projectors.sec}, we will explicitly construct the diffuson projectors
onto irreducible subspaces. For spin $\frac{1}{2}$, this reduces indeed to the electronic projectors
\refeq{calT0.electron} and  \refeq{calT1.electron}.
The cooperon vertex projectors \refeq{T0.electron} and  \refeq{T1.electron}
will be shown to be given by
\begin{eqnarray}
    T^{(0)}\abcd  & = & - \frac{1}{2}
    \calT^{(0)}_{\alpha\delta,\gamma\beta}
    + \frac{1}{2}  \calT^{(1)}_{\alpha\delta,\gamma\beta}\ , \\
T^{(1)}\abcd & = &\frac{3}{2}
    \calT^{(0)}_{\alpha\delta,\gamma\beta}
    + \frac{1}{2}\calT^{(1)}_{\alpha\delta,\gamma\beta}\ .
\end{eqnarray}
This is in fact the spin $\frac{1}{2}$ version of the more general
relation 
\eqlab{ 
T^{(K)}\abcd  =  \sum_{K'} R_s(K,K')
        \, \calT^{(K')}_{\alpha\delta,\gamma\beta}
}{transfbase} 
between the cooperon and diffuson projectors, valid
for arbitrary spin, to be derived in section
\ref{reltensprop.sec}. Here, our notation 
\eqlab{ R_s(K,K')=
(2K+1) \sixj{s}{s}{K}{s}{s}{K'} }
{Rs} 
is a $6j$-symbol from
standard angular momentum theory \cite{Edmonds,Messiah}. Thanks to the
$6j$-symbol orthogonality \cite[(35$c$)]{Messiah}
\footnote{Instead of compiling a large appendix,
we will refer to standard definitions and sum rules by citing the
exact location in appendix C of Messiah's book, e.g., his equation
(35$a$) by writing \cite[(35$a$)]{Messiah}.} 
\begin{equation}\label{orthoRs}
\sum_{K'} R_s(K,K') \, R_s(K',K'') = \delta_{K,K''},  
\end{equation}
the inverse relation to \refeq{transfbase} is equally simple: 
\begin{equation}\label{basetransf.eq}
\calT^{(K)}\abcd
= \sum_{K'} R_s(K,K') \, T^{(K')}_{\alpha\delta,\gamma\beta}\ .
\end{equation}

\subsubsection{Eigenvalues}

We will show in section \ref{eigenvalues.sec} that the eigenvalues of the normalized spin-flip vertices
\refeq{defV.spin-flip} and \refeq{defX.spin-flip} are given by
\begin{equation}
    \lambda_K  = 1 - \frac{K(K+1)}{2s(s+1)} = - \chi_K.
\end{equation}
The eigenvalues for an arbitrary microscopic spin scattering
potential will be derived below in full generality. In all cases,
the eigenvalues of a scalar scattering vertex are linked by the
recoupling relations 
\eq{ \lambda_K = \sum_{K'} R_s(K',K)\chi_{K'}, 
\quad \quad \chi_{K} = \sum_{K''} R_s(K'',K)
\lambda_{K''}. } 
These relations between eigenvalues take the form of a contravariant
transformation of coordinates associated with the respective covariant transformation
\refeq{transfbase} and \refeq{basetransf} of  the projectors.  

These results should  motivate our readers to consider with interest the
following, more involved derivations. In the following section \ref{projectors.sec}, we lay the algebraic
foundations of the decomposition by deriving the orthogonal
projectors, before turning to the eigenvalues in section
\ref{eigenvalues.sec}.

\section{Diagonalization of intensity vertices}
\label{projectors.sec}

Let us now consider a general spin interaction defined by its matrix elements
$\bra{s\gamma}V\ket{s\alpha}$ for arbitrary spin $s$.
The corresponding diffuson and cooperon vertices are given by
\begin{eqnarray}
 \calL\abcd = \mv{ \bra{s\gamma}V\ket{s\alpha}  \bra{s\beta} V^\dagger
\ket{s\delta}} \ , 
\label{defL.general.eq} \\ 
X_{\alpha\beta,\gamma\delta} = \calL_{\alpha \delta,\gamma\beta}
= \mv{
\bra{ s\gamma } V \ket{s \alpha} \bra{s \delta }V^\dagger
\ket{s\beta}} \ .\label{defX.general.eq}
\end{eqnarray}

\subsection{The diffuson vertex as a spin superoperator}

Spin states $\ket{s\alpha}$ are vectors in the
Hilbert space  $H_s=\mathbbm{C}^{d_s}$ with
dimension $d_s=2s+1$.
The diffuson vertex $\calL$ is a
linear mapping $\ket{s\alpha}\bra{s\gamma}\mapsto
\ket{s\delta}\bra{s\beta}$  between spin operators.
Its argument space therefore is the space of linear operators acting on
$H_s$, the so-called Liouville space $L(H_s)$ with dimension $d_s^2$
\cite{Blum96}. Any linear operator
$A\in L(H_s)$ is simply a $d_s\times d_s$ matrix. The
trace-preserving vertex $\calL: L(H_s) \to L(H_s)$ then is a
\textit{superoperator} (thus the notation with a curly script),
mapping a matrix $A$ onto another matrix $A' = \mathcal{L}A$. Its
action in the basis of spin projectors $\{ \ket{s\alpha}\bra{s\beta}\}$
reads $A'_{\gamma\delta} =
\calL\abcd A_{\alpha\beta}$ in terms of the matrix elements
\eqlab{
    \calL\abcd =    \tr\left\{(\ket{s\delta}\bra{s\gamma}) \; \calL \;
(\ket{s\alpha}\bra{s\beta})\right\}\ . }
{Vabcd}
Here $\tr\{\cdot\}=\sum_\alpha\bra{s\alpha} \cdot \ket{s\alpha}$ is the trace over $H_s$. In superoperator
notation, the diffuson vertex reads 
\eqlab{
    \calL =  \sum_{\alpha\beta\gamma\delta} (\ket{s\gamma}\bra{s\delta})
\calL\abcd \; \tr\left\{(\ket{s\beta}\bra{s\alpha}) \ \cdot\
\right\} }{Vop.trace} 
With this notation, the ressemblance with
the Liouvillian $\calL = -\frac{i}{\hbar} [H,\ \cdot\ ]$, the
generator of time evolution, becomes apparent. 
We define the trace of $\calL$ as a linear operator in the Liouville space
as 
\begin{equation}\label{traceL.eq}
\mathrm{Tr_L} \calL = \sum_{\alpha,\beta}
\calL_{\alpha\beta,\,\alpha\beta}\ .
\end{equation}

In the extensive
literature on Liouville space formalism \cite{laloe,
CCT,gabriel,omont73}, one often views the spin operators as vectors in
$L(H_s)$ and defines corresponding kets by 
$\ket{s\beta}\bra{s\alpha} = \ket{\alpha \beta}\rangle$. Using
this notation, the diffuson vertex matrix elements are given by
    $\calL\abcd =    \langle\langle  \gamma\delta | \calL | \alpha\beta \rangle\rangle$,
and the superoperator takes the very simple form
$\calL =  \sum_{\alpha\beta\gamma\delta}   \calL\abcd |\gamma\delta
    \rangle\rangle \ \langle\langle \alpha\beta |$.
In this notation, the parallel with the cooperon vertex operator
(see \refeq{Xopdecouple} below) is especially clear.
However, we deliberately choose to use the
superoperator formulation in the following
because it allows us in section \ref{generators.sec}
to derive the ladder vertex projectors in terms of spin operators
(which is needed to get expression \refeq{calT1.electron}).
Moreover, to answer completely question 1 raised in section \ref{heur.diag.sec},
we have to explain the \textit{difference} between the diffuson and
cooperon eigenstructures rather than their similarity.
This difference reflects the different behaviour under rotations of spin states $\ket{s\alpha}$ and
their conjugates $\bra{s\alpha}$ that are explicitly featured in the superoperator notation \refeq{Vop.trace}.

\subsection{Decomposition into irreducible superoperators}

An incident state $\ket{s\alpha}$ in scattering amplitudes like
\refeq{Vm}
is a spinor, i.e.,  a vector in $H_s$
that
transforms under the irreducible representation $D^{(s)}$
of the rotation group SU(2):
$   \ket{s\alpha}\mapsto U\ket{s\alpha}
    =\sum_\mu \ket{s\mu}\bra{s\mu}U\ket{s\alpha}= \sum_\mu U_{\mu\alpha} \ket{s\mu}
$ with an appropriate unitary and unimodular matrix
$UU^\dagger=\mathbbm{1}_{d_s}$, $\det U=+1$. A final state
$\bra{s\gamma}$, however, transforms contragradiently \cite{FR59},
i.e. , under the complex conjugate representation
$\overline{D^{(s)}}$, $
    \bra{s\gamma}\mapsto \bra{s\gamma}U^\dagger = \sum_\nu
    \overline{U}_{\nu\gamma} \bra{s\nu}
$.
Under a rotation, the complete spin vertex is transformed as
$    \calL_{\alpha\beta,\gamma\delta} \mapsto
    U_{\sigma\delta}(U^\dagger)_{\gamma\rho}
    \calL_{\mu\nu,\rho\sigma} U_{\mu\alpha}
    (U^\dagger)_{\beta\nu}$.
Clearly, this vertex is not a rank four tensor (that would
transform under the direct product $(\mathcal{D}^{(s)})^{\otimes
4}$), but rather a two-by-two mixed tensor that transforms under
$(D^{(s)}\otimes\overline{D^{(s)}})^{\otimes 2}$.
If the vertex is a scalar, it is invariant under this transformation.

In expressions \refeq{Vabcd} and \refeq{Vop.trace},
 the operator arguments of $\calL$ are decomposed
over the decoupled product basis $\{\ket{s\alpha}\bra{s\beta}\}$ of the
Liouville space
$L(H_s)$. But according to our  diagonalization strategy, we want
 to use a basis adapted to irreducible representations
of the rotation group.
We first perform the  Clebsch-Gordan (CG in short) decomposition of the argument and
image representations $D^{(s)}\otimes\overline{D^{(s)}}$.
These  are then recoupled in turn to give the complete CG-decomposition of
$(D^{(s)}\otimes\overline{D^{(s)}})^{\otimes 2}$ which yields the irreducible components of the superoperator.
The route thus taken may be traced in the following map:
\eqlab{
\begin{array}{ccc}
(D^{(s)} \otimes \overline{D^{(s)}}) &   &  (D^{(s)}  \otimes
\overline{D^{(s)}})\\
  \searrow \quad  \swarrow & &   \searrow \quad  \swarrow  \\
 D^{(K)}  &\otimes &  D^{(K')} \\
 \qquad \qquad    \searrow &   & \swarrow \qquad  \qquad \\
 & D^{(L)}&
\end{array}
}{CGroute}
Following standard procedures from angular momentum theory
\cite{FR59,Blum96}, one can define a set of irreducible
operators adapted to our purpose,
\eqlab{
    T^{(K)}_q = T^{(K)}_q(s,s)=\sum_{mm'}(-)^{s-m}\cg{ssm'{-m}}{Kq}\
    \ket{sm'}\bra{sm}
}{TKq}
with matrix elements
\eqlab{
\bra{sm'}T^{(K)}_q \ket{sm} =
(-)^{s-m}\cg{ssm'{-m}}{Kq}
}
{TKq.mm'}
where $\cg{ssm'{-m}}{Kq}$ are the usual CG-coefficients.
These types of tensors,
called ``statistical tensors'' or ``state multipoles'' are
irreducible components of the density matrix and have been
developed by U.\ Fano and G.\ Racah in the 1950s
\cite{FR59}. Their
construction is very similar to the coupling scheme of angular
momentum eigenstates: one simply chooses a linear combination of
spin projectors with suitable CG-coefficients. The definition
\refeq{TKq} features a characteristic minus sign  in front of the
spin quantum number $m$ that is
reminiscent of the contragradient transformation of $\bra{sm}$.
 Hermitian conjugation is defined by
$T^{(K)\dagger}_q= (-)^q T^{(K)}_{-q}$.

The orthogonality of CG-coefficients assures that the operators
\refeq{TKq} are orthonormalized with respect to the  matrix
scalar product $(A|B) = \tr\{A^\dagger B\}$,
 \eqlab{
 \psop{T^{(K)}_q}{T^{(K')}_{q'}} =
\tr \left\{ T^{(K)\dagger}_q  T^{(K')}_{q'}\right\}
  = \delta_{KK'}\delta_{qq'} \ .
}{diagTKq}
The set of irreducible tensor operators $T^{(K)}_q$ provides a
natural basis that incorporates best the rotational symmetries.
Any linear operator $O$ can be developed in this basis according
to $O=\sum_{Kq} O_{Kq} T^{(K)}_q$, with components
\eq{ O_{Kq} = \psop{T^{(K)}_q}{O}=
\sum_{mm'}(-)^{s-m}\cg{ssm'{-m}}{Kq}\bra{sm'} O \ket{sm}. }
Inserting
$\ket{sm'}\bra{sm} =
\sum_{Kq} (-)^{s-m} \cg{ssm'{-m}}{Kq}
    T^{(K)}_q $
in the vertex definition \refeq{Vop.trace}, the
superoperator becomes
\eq{
    \calL = \sum_{KqK'q'}  T^{(K')}_{q'} \calL^{(K)}_{q,K'q'}
        \psopdot{T^{(K)}_{q}}
}
where its left-right irreducible components are $\calL^{(K)}_{q,K'q'} = \psop{T^{(K')}_{q'}}{ \calL T^{(K)}_q}$ or
\eqlab{
\fl
\calL^{(K)}_{q,K'q'}   =  \sum_{\alpha\beta\gamma\delta}
    (-)^{s-\beta} \cg{ss\alpha\,{-\beta}}{Kq} \; \calL\abcd \;  (-)^{s-\delta}
\cg{ss\delta\,{-\gamma}}{K'{-q'}} .
}{VKq}
Now we recouple the irreducible argument and image representations to
get the complete CG-decomposition (last line in \refeq{CGroute}).   
We define a basis of irreducible superoperators
$\calT^{(L)}_m(K,K')$ of rank $L$ with components $m=-L,\dots,L$:
\eq{
\calT^{(L)}_m(K,K') = \sum_{qq'} (-)^{K-q}\cg{K'Kq'{-q}}{Lm} T^{(K')}_{q'}
\psopdot{T^{(K)}_{q}}
}
The recoupled objects can be precisely located on the decomposition map \refeq{CGroute}:
\eqlab{
\begin{array}{ccc}
 \ket{s\alpha}\ \bra{s\beta} & &  \ket{s\delta}\ \bra{s\gamma}  \\
   \searrow \ \swarrow  & &    \searrow \ \swarrow \\
 T^{(K)}_q  &  &  T^{(K')}_{q'} \\
 \qquad \qquad    \searrow &   & \swarrow \qquad  \qquad \\
 & \calT^{(L)}_m (K,K')&
\end{array}
}{}
The
vertex in irreducible superoperator notation is then
\eqlab{
\calL = \sum_{KK'}\sum_{Lm} \calL_{Lm}(K,K') \calT^{(L)}_m(K,K')
}{VLmdef}
with components
$\calL_{Lm}(K,K') = \sum_{qq'} (-)^{K-q}
\cg{K'Kq'{-q}}{Lm} \calL^{(K)}_{q,K'q'}
$.

\subsection{Scalar diffuson vertex}

The above basis set construction and
decomposition into irreducible superoperators
apply to arbitrary superoperators.
This basis change is especially profitable
when the diffuson vertex under consideration is a
scalar with respect to rotations. In this case, its only non-vanishing irreducible component
is $\calL_{00}(K,K')$ for $L=0,m=0$. The usual
selection rules of CG-coefficients then require in \refeq{VLmdef} that $K=K'$ and
$q=q'$: scalar superoperators indeed connect
irreducible subspaces $L(H_s)^{(K)}$ with equal rank $K$. Each of
these subspaces has dimension ($2K+1$) and the total dimension is
of course preserved, $\sum_{K=0}^{2s} (2K+1) = d_s^2 = (2s+1)^2$.
As found heuristically in section \ref{heur.diag.sec},
$\calL$ then takes the form
\eqlab{
    \calL = \sum_{K=0}^{2s} \calL_{00}(K,K) \calT^{(0)}_0(K,K)
= \sum_{K=0}^{2s} \lambda_K  \calT^{(K)}
}{Vdiagonal}
with $\lambda_K = \calL_{00}(K,K)/\sqrt{2K+1} $ and
$\calT^{(K)} = \sqrt{2K+1} \; \calT^{(0)}_0(K,K)$.
The calculation of eigenvalues will be treated in detail in section \ref{eigenvalues.sec}.
We now complete the algebraic characterization of the projectors.

\subsubsection{Properties of the scalar projectors}
\label{diffusonprojectors.sec}

The superoperators $\calT^{(K)}$ which diagonalize a scalar vertex
are projectors onto the Liouville subspaces
$L(H_s)^{(K)}$ of irreducible operators of rank $K$:
\eqlab{
\fl
 \calT^{(K)} = \sqrt{2K+1}\; \calT^{(0)}_0(K) = \sum_q  T^{(K)}_{q}
\psopdot{T^{(K)}_{q}}
=  \sum_q  T^{(K)}_{q} \tr\left\{T^{(K)\dagger}_{q}\  \cdot\ \right\}
}{calTK}
These operators are scalar objects themselves
since $\sum_q T^{(K)}_q T^{(K)\dagger}_q$ generalizes the scalar
product between vector operators ($K=1$) to arbitrary rank $K$
\cite[(87)]{Messiah}.
Thanks to the orthogonality  \refeq{diagTKq} of the basis tensors,
the $\calT^{(K)}$'s are indeed orthogonal projectors, 
\eq{
\calT^{(K)} \calT^{(K')} = \sum_{qq'}  T^{(K)}_{q}
\underbrace{    \psop{T^{(K)}_q}{T^{(K')}_{q'}}
}_{\delta_{KK'}\delta_{qq'}}
\psopdot{T^{(K')}_{q'}}
= \delta_{KK'} \calT^{(K)}\ .
}
Their matrix elements in the decoupled basis
$\{\ket{s\alpha}\bra{s\gamma}\}$  of spin projectors
are found by inserting \refeq{calTK} into the superoperator
definition  \refeq{Vabcd},
\eqlab{
 \calT^{(K)}_{\alpha\beta,\gamma\delta}  =
\tr \left\{(\ket{\delta}\bra{\gamma})\calT^{(K)}
(\ket{\alpha}\bra{\beta})\right\} = \sum_q \bra{\gamma}T^{(K)}_{q} \ket{\delta}
\bra{\beta}T^{(K)}_{q}{}^\dagger \ket{\alpha}\ .
}{TKqabcd} 
Using the matrix elements \refeq{TKq.mm'} and the completeness
relation of CG-coefficients \cite[(14$a$)]{Messiah}, it is
straightforward to show that 
\begin{equation}\label{TrProL}
\mathrm{Tr_L}\calT^{(K)} = \sum_{\alpha\beta}
\calT^{(K)}_{\alpha\beta,\,\alpha\beta} = 2K+1,
\end{equation}
as expected for the identity in the subspace $L(H_s)^{(K)}$ of
dimension $2K+1$.  
Furthermore, the projectors
$\calT^{(K)}$ sum up to the identity 
with respect to the horizontal  product rule \refeq{defproduit}:   
$\sum_K
\calT^{(K)}_{\alpha\beta,\gamma\delta} = I \abcd
= \delta_{\gamma\alpha}\delta_{\beta\delta}$. 

\subsubsection{Expression in terms of spin operators}
\label{generators.sec}

The projector onto scalar operators is the ``trace-taker''
\eqlab{
    \calT^{(0)}
    = \frac{1}{d_s}\ \mathbbm{1} (\mathbbm{1},\ \cdot\ )
    = \frac{1}{d_s}\ \mathbbm{1}\ \tr\{\ \cdot\ \} \ ,
}{calT0}
with $\mathbbm{1}$ the identity in $H_s$,  
which is all but a surprise considering that the scalar
part of a matrix is its trace. In the decoupled basis, we have
$    \calT^{(0)}_{\alpha\beta,\gamma\delta}
    = \frac{1}{d_s}\delta_{\alpha\beta}\delta_{\gamma\delta}$.
justifying thereby the tensor \refeq{calT0.electron}
found by elementary diagonalization in the electron case. Now it
is evident that a unit superoperator eigenvalue $\lambda_0=1$ is
equivalent with trace-preservation (and hence particle/energy
conservation).

But already for the projector $ \calT^{(1)}$ onto vector operators,
using  \refeq{TKqabcd} involves 
a sum over products of CG-coefficients, and it is advisable to look
for a more transparent formulation.  Equation \refeq{calTK} tells
us that we need to find a contraction $\calT^{(1)}= \sum_j O^j
\psopdot{O^j}$  of components of a vector operator $\bi
O=(O^1,O^2,O^3)$ that must be traceless, $\sum_\alpha
O^j_{\alpha\alpha}=0$, in order to be orthogonal to $\calT^{(0)}$.
The only available vector is the generator of rotations: the spin
operator $\bi S$ itself. Its components have zero trace because
they generate rotation matrices of unit determinant ($1=\det U =
\det \exp\{i \theta S^j\}= \exp\{i \theta \tr S^j\}$). 
Alternatively, one can use the Wigner-Eckart theorem
\cite[(84)]{Messiah} to show explicitly 
that $S_q = \sqrt{c_s} T^{(1)}_q$ up to a normalisation constant such
that
\eqlab{
\calT^{(1)}= \frac{1}{c_s} \sum_j S^j
\psopdot{S^j}. }
{calT1}
The normalisation constant $c_s$ is fixed
by requiring $\calT^{(1)}\calT^{(1)}=\calT^{(1)}$: Since $S^2 =
s(s+1)\mathbbm{1}_s$ is the Casimir operator of the irreducible
representation $D^{(s)}$ of dimension $d_s=2s+1$, we have
$3\,\tr\left\{S^i S^j\right\} = s(s+1)(2s+1)\, \delta_{ij} $. The
normalization factor therefore is
$c_s = s(s+1)(2s+1)/3$.
For electrons, $S^j= \sigma^j/2$ and
$c_{1/2}=\frac{1}{2}$ such that $\calT^{(1)} = \frac{1}{2}\sum_j \sigma^j
(\sigma^j, \ \cdot \ )$. Its components in the decoupled basis are
indeed those of \refeq{calT1.electron}.

This completes the derivation of projectors for the scalar electronic
diffuson vertex. For larger spin, higher orders of $K$ have to be
considered which essentially involves a Gram-Schmidt procedure. 
In \ref{gram-schmidt.sec}, this is done for the photon
case $s=1$.

\subsection{The crossed vertex as an ordinary operator}

We now turn to the diagonalization of the cooperon vertex $X$ that
maps an incident tensorial ket
product $|s\alpha \rangle \otimes |s\beta \rangle$ onto the
final tensorial ket product $|s\gamma \rangle \otimes |s\delta
\rangle$.
In operator form,
\eqlab{
    X = \sum\abcd \ket{s\gamma\, s\delta} X \abcd \bra {s\alpha\,
s\beta}\ .
}{Xopdecouple}
Therefore, the crossed vertex $X: H_s\otimes H_s \to H_s\otimes H_s$
can be seen as an ordinary linear operator or $d_s^2\times d_s^2$
matrix. In this respect, we define the trace for $X$ as:
\eqlab{
\mathrm{Tr_C}X = \sum_{\alpha\beta} X_{\alpha\beta,\,\alpha\beta}.
}{traceC}
This trace definition is invariant under the exchange of spin indices
\refeq{defX.general}, such that 
\begin{equation}\label{traceconserv}
\mathrm{Tr_L}\calL = \mathrm{Tr_C}X
\end{equation}
In other words, 
the partial Liouville conjugation \cite{Ben-Reuven66}  $\ket{s\beta}\bra{s\delta}
\mapsto \ket{s\delta}\bra{s\beta}$ that maps the
diffuson onto the cooperon vertex preserves  
their trace, which in turn will permit us to derive useful sum
rules between eigenvalues in section \ref{rel.eigenvalues.sec}. 

We wish to bring the vertex into a least redundant form for iteration
by performing the usual Clebsch-Gordan decomposition of the argument
and image spaces, 
i.e., a suitable basis change that transforms the direct
product $D^{(s)}\otimes D^{(s)}$ of two irreducible
representations acting on $H_s\otimes H_s$
 into the direct sum $D^{(0)}\oplus D^{(1)} \oplus
\dots \oplus D^{(2s)}$ of  irreducible representations $ D^{(K)}$,
$K=0,1,\dots,2s$.
The appropriate recoupling route now can be  mapped out as
\eqlab{
\begin{array}{ccc}
(D^{(s)} \otimes D^{(s)})  & &  (\overline{D^{(s)}}  \otimes
\overline{D^{(s)}})  \\
\searrow\quad \swarrow  & &  \searrow\quad \swarrow   \\
D^{(K)}  &\otimes &   \overline{D^{(K')}} \\
 \qquad \qquad    \searrow &   & \swarrow \qquad  \qquad \\
 & D^{(L)}&
\end{array}
}{CGrouteX}
We first change to the spherical basis $\ket{Kq} = \sum_{\alpha\beta} \cg{ss\alpha\beta}{Kq}
\ket{s\alpha\, s\beta}$ of the irreducible subspace
$H^{(K)}$.
The corresponding vertex components are
\eq{
X_{Kq,K'q'} =\sum\abcd
\cg{ss\alpha\beta}{Kq} X \abcd \cg{ss\gamma\delta}{K'q'}
}
such that
$X = \sum_{Kq,K'q'} \ket{Kq}X_{Kq,K'q'} \bra {K'q'}$.
Here, $X$ is decomposed over the decoupled  operator basis $\ket{Kq} \bra {K'q'}$.
We can therefore define recoupled irreducible operators, as in \refeq{TKq}, but in the present context with
rank $L$ and their ($2L+1$) components
\eqlab{
T^{(L)}_m(K,K') = \sum_{qq'} (-)^{K'-q'} \cg{KK'q\,
{-q'}}{Lm}\; \ket{Kq} \bra {K'q'}\ . }
{TLm}
The objects thus used can be precisely located on the map
\refeq{CGrouteX},
\eqlab{
\begin{array}{ccc}
 \ket{s\alpha}\ket{s\beta} & &  \bra{s\delta} \bra{s\gamma}  \\
 \searrow\  \swarrow  & &    \searrow\  \swarrow \\
  \ket{Kq}   &  &  \bra{K'q'} \\
 \qquad \qquad    \searrow &   & \swarrow \qquad  \qquad \\
 & T^{(L)}_m (K,K')&
\end{array}
}{Xobjects}
The $T^{(L)}_m(K,K')$'s provide the most
natural basis set for exploiting rotational symmetries in $H_s
\otimes H_s$. Now the crossed vertex $X$ can be decomposed over
this basis set: 
\eqlab{
 X = \sum_{KK'} \sum_{Lm} X_{Lm}(K,K')  T^{(L)}_m(K,K')
}{XLmdef}
where its irreducible components are
\eqlab{
X_{Lm}(K,K')  = \sum_{qq'} (-)^{K'-q'} \cg{KK'q\, {-q'}}{Lm}\;
X_{Kq,K'q'}\ .
}{XLm}

\subsection{Scalar crossed vertex: projectors}
\label{cooperonprojectors.sec} 
The decomposition into ireducible components is especially profitable
for a scalar vertex since it has only a single non-vanishing component
$X_{00}(K,K)$. The CG-coefficients in \refeq{TLm} and \refeq{XLm}
then restrict the sum to $K=K'$. Therefore, the vertex connects
irreducible sub-spaces $H^{(K)}$ of equal rank and can indeed be
written as
\eqlab{
X = \sum_{K=0}^{2s} \chi_K   T^{(K)}
}{Xdiag}
with eigenvalues $\chi_K  = X_{00}(K)/\sqrt{2K+1}$ and
associated tensors
\eqlab{
T^{(K)} = \sqrt{2K+1} \; T^{(0)}_0(K) = \sum_q
\ket{Kq}\bra{Kq}.
}{TK}
These are indeed orthogonal projectors
$
T^{(K)}T^{(K')} = \delta_{KK'}T^{(K)}
$
onto the irreducible
subspaces $H^{(K)}$. Their matrix elements
in the decoupled basis of $H_s\otimes H_s$ are
\eqlab{
    T^{(K)}\abcd = \sum_q \cg{ss\alpha\beta}{Kq}
        \cg{ss\gamma\delta}{Kq},
}{TKqabcd.cooperon} 
Using \cite[(14$a$)]{Messiah}, one shows that
\begin{equation}\label{TrProX}
\mathrm{Tr_C}T^{(K)} = \sum_{\alpha\beta}
T^{(K)}_{\alpha\beta,\,\alpha\beta} = 2K+1
\end{equation}
and that the projectors $T^{(K)}$ sum up to the identity, 
$\sum_K
T^{(K)}\abcd = \delta_{\gamma\alpha}\delta_{\beta\delta}$
.

Even the simplest projector $T^{(0)}$ on the singlet space
$H^{(0)}$ has 
seemingly complicated
matrix elements in the decoupled basis,
\eqlab{
    T^{(0)}\abcd = \frac{1}{d_s} (-)^{2s+\beta-\gamma}
        \delta_{-\beta,\alpha}\delta_{-\delta,\gamma}\ .
}{T0abcd}
For electrons, one may check by hand that this is
indeed equivalent to the much nicer formula \refeq{T0.electron}.
For spin 1 particles,
the contractions $(-)^p\delta_{-p,q}$ of spherical basis
components become $\delta_{ij}$ in the cartesian basis, and one
gets \eq{
    T^{(0)}_{il,jk} = \frac{1}{3} \delta_{il}\delta_{jk}
}
as used in
\cite{Mueller02}.
These heuristic writings in the decoupled basis are much less
systematic than the exceedingly simple form \refeq{TK} in the spherical
basis. Moreover, the matrix elements of $T^{(1)}$ 
directly derived from \refeq{TKqabcd.cooperon} 
would be rather
far from the simple form \refeq{T1.electron} we
wish to justify. To that purpose, section \ref{reltensprop.sec} discusses a general way of deriving the crossed
projectors $T^{(K)}$ from the ladder (super-)projectors $\calT^{(K)}$
and vice versa.

\subsection{Recoupling of projectors}
\label{reltensprop.sec}

The simple exchange rule $\calL\abcd = X_{\alpha\delta,\gamma\beta}$
provides us with a convenient way of linking the diffuson and cooperon projectors.
The two diagonalization procedures differ by the coupling scheme for the two pairs of spin indices
and are related by a simple recoupling relation.
Indeed, the matrix elements \refeq{TKqabcd} of the diffuson projectors,
\eq{
\calT^{(K)}_{\alpha\beta,\gamma\delta}  =  \sum_q (-)^{s-\delta} \cg{ss\gamma\,{-\delta}}{Kq}   (-)^{s-\beta}
\cg{ss\alpha\,{-\!\beta}}{Kq}
}
can be derived from the corresponding cooperon projectors \refeq{TKqabcd.cooperon} with exchanged spin
indices $\beta\leftrightarrow \delta$,
$T^{(K)}_{\alpha\delta,\gamma\beta}  = \sum_q \cg{ss\alpha\delta}{Kq}
        \cg{ss\gamma\beta}{Kq}$,
with the help of the appropriate recoupling relation
\cite[(34)]{Messiah}: 
\eqlab{
    T^{(K)}\abcd  = \sum_{K'}
        R_s(K,K') \, \calT^{(K')}_{\alpha\delta,\gamma\beta}, \quad 
 \calT^{(K)}\abcd = \sum_{K'} R_s(K,K')
        \, T^{(K')}_{\alpha\delta,\gamma\beta}\ .
}{TKdecalTK'} 
These relations express a 
mapping between sets of projector matrix elements $\{\calT^{(K)}\abcd\} \leftrightarrow
\{T^{(K)}_{\alpha\delta,\gamma\beta}\}$  defined by a 
transformation matrix $R_s$ with elements 
\eq{ R_s(K,K')=
(2K+1) \sixj{s}{s}{K}{s}{s}{K'}\ .} 
General  $6j$-symbol symmetry properties and the orthogonality
(\ref{orthoRs}) imply that the 
matrix $R_s$ is real and circular: $ R_s = \overline{R_s} =
R_s^{-1}$ such that $\det R_s=\pm1$ with $\det R_{1/2}=\det R_1=-1$. 
This transformation conserves the orthogonality of projectors:  
$
\Tr_\mathrm{L}\{ \calT^{(K)}\calT^{(K')} \} = (2K+1)\delta_{KK'} = \Tr_\mathrm{C}\{T^{(K)}
T^{(K')} \}$.  

Putting the transformations \refeq{TKdecalTK'} to work, 
the scalar projector \refeq{T0abcd} on the singlet state
$K=0$ is predicted to be given by 
\eq{
    T^{(0)}\abcd  = \frac{(-)^{2s}}{d_s} \sum_{K'} (-)^{K'}
    \calT^{(K')}_{\alpha\delta,\gamma\beta}\ . 
}
Using the electron diffuson projectors \refeq{calT0.electron} et
\refeq{calT1.electron}, one obtains as expected the singlet cooperon 
projector in the form \refeq{T0.electron},
\eq{
    T^{(0)}\abcd  = - \frac{1}{2} \left (
    \calT^{(0)}_{\alpha\delta,\gamma\beta}
    - \calT^{(1)}_{\alpha\delta,\gamma\beta}\right )
        = \frac{1}{2} \left
        ( \delta_{\alpha\gamma}\delta_{\beta\delta}
        - \delta_{\alpha\delta}\delta_{\beta\gamma} \right)\ .
}
Similarly, the projector \refeq{T1.electron} onto the triplet space is
\eq{
    T^{(1)}\abcd  = 3 \left ( \frac{1}{2}
    \calT^{(0)}_{\alpha\delta,\gamma\beta}
    + \frac{1}{6}\calT^{(1)}_{\alpha\delta,\gamma\beta}\right )
        = \frac{1}{2} \left
        ( \delta_{\alpha\gamma}\delta_{\beta\delta}
        + \delta_{\alpha\delta}\delta_{\beta\gamma} \right)\ .
}
This completes the derivation of all projectors for the case of spin $\frac{1}{2}$.
The photon case is discussed in \ref{photon.sec}.
Now the stage is set for the calculation of eigenvalues.

\section{Calculation of eigenvalues}
\label{eigenvalues.sec}

\subsection{Diffuson eigenvalues $\lambda_K$}
\label{laddereigenvalues.sec}


The scalar spin superoperator eigenvalues $\lambda_K$ defined through the general decomposition
\refeq{Vdiagonal} are at least ($2K+1$)-fold
degenerate.
\footnote{Of course, larger degeneracies occur if the vertex possesses even
higher symmetries; an elementary example is the identity
$I\abcd = \delta_{\gamma\alpha}\delta_{\delta\beta}$
with its single $d_s^2$-fold degenerate eigenvalue $\lambda=1$.
} 
They can be calculated either by
projecting the vertex on an arbitrary component $q$ of the
respective subspace, $\lambda_K = \psop{T^{(K)}_q}{\calL
T^{(K)}_q}$, or directly from the vertex matrix elements  $\calL\abcd$
as
\begin{eqnarray}
\fl \lambda_K =  \frac{\calL_{00}(K,K)}{\sqrt{2K+1}}
        = \frac{1}{2K+1} \sum_q \calL^{(K)}_{q,Kq}  \\
 \lo  =  \frac{1}{2K+1} \sum_{q,\alpha\beta\gamma\delta}
    (-)^{s-\beta} \cg{ss\alpha\,{-\beta}}{Kq}\; \calL\abcd \;  (-)^{s-\delta}
\cg{ss\delta\,{-\gamma}}{K{-q}}.
\label{lambdaK.eq}
\end{eqnarray}
Useful information about the possible form of
eigenvalues can be gained from this direct calculation. 
To that purpose, consider an arbitrary  microscopic spin interaction with matrix
elements $V_{\gamma\alpha}=\bra{s\gamma}V\ket{s\alpha}$. This
interaction can itself be developed in the basis of irreducible
operators \refeq{TKq}, $V  = \sum_{Kq} V_{Kq} T^{(K)}_q$. Its
components
\eqlab{
V_{Kq} = \psop{T^{(K)}_q}{V} =
\tr\left\{T^{(K)\dagger}_q V\right\} =
 \sum_{\alpha\gamma} (-)^{s-\alpha}\cg{ss\gamma\,{-\alpha}}{Kq}\, V_{\gamma\alpha}
}{VmicroKq}
are the coupling amplitudes of scalar, vector,
quadrupolar type, etc. These amplitudes may depend on microscopic
degrees of freedom of the scattering object such as the
orientation of a magnetic impurity. The scattering vertex
$\calL\abcd = \mv{V_{\gamma\alpha} (V^\dagger)_{\beta\delta}}$ is
the partial trace over these degrees of freedom,
\eq{\fl 
\calL\abcd =
\sum_{KqK'q'} \mv{V_{Kq}\overline{V_{K'q'}}} (-)^{2s-\alpha-\beta}
\cg{ss\gamma\,{-\alpha}}{Kq} \cg{ss\beta\,{-\delta}}{K'{-q'}}
}
As a scalar vertex has no angular dependence, it  always take the
generic form
\eqlab{
 \mv{V_{Kq}\overline{V_{K'q'}}} = \delta_{KK'}\delta_{qq'}\,s_K
  }{def.sK}
where $s_K$ is the vertex eigenvalue in the invariant subspace of
rank $K$ for the ``vertical'' coupling scheme. 
The eigenvalue definition \refeq{lambdaK} then becomes a sum over
products of four CG-coefficients that defines a $6j$-symbol
\cite[(32)]{Messiah} and finally yields: 
\eqlab{ \lambda_K =
(-)^{2s+K} \sum_{K'}(-)^{K'} \, R_s(K',K) \, s_{K'} \ .
}{lambdaK.sK} 
Remarkably enough, the only information about the microscopic
interaction is carried  by the coupling constants $s_{K}$. The
$6j$-symbol merely provides the recoupling from the ``vertical''
form $(\alpha\gamma)\leftrightarrow(\beta\delta)$ of the initial
product of amplitudes
$\mv{V_{\gamma\alpha}(V^\dagger)_{\beta\delta}}$ to the
``horizontal'' form $(\alpha\beta)\leftrightarrow(\gamma\delta)$
necessary for the diagonalization with respect to the multiple
scattering product rule \refeq{defproduit}.

The simplest possible example of a scalar diffuson vertex is of
course given by isotropic scattering. The interaction is characterized
by $V^{(0)}_{\gamma\alpha} = v_0 \delta_{\gamma\alpha}$ with
$v_0$ being a complex number. The only effective amplitude is then $s_0 =
(2s+1) |v_0|^2$, and the eigenvalues are simply $\lambda_K =
|v_0|^2$ for all $K$ which is evident at once
since this vertex is simply proportional to the identity.

The first non-trivial example is given by a vector coupling of the
spin-flip form (\ref{defV0.spin-flip.eq}). 
Using the chosen normalization \refeq{defV.spin-flip} and $S_q =
\sqrt{s(s+1)d_s/3}  \; T^{(1)}_q$ by virtue of the
Wigner-Eckart theorem \cite[(84)]{Messiah}, one finds that the only
nonzero coupling is the vector contribution $s_1 = d_s/3$. Using
\refeq{Rs}, the eigenvalues \refeq{lambdaK} are then
\eqlab{
    \lambda_K =  (-)^{d_s+K} d_s \, \sixj{s}{s}{K}{s}{s}{1}
= 1 - \frac{K(K+1)}{2s(s+1)}.
} {lambdaK_spin-flip_6j}
The scalar eigenvalue is identically $\lambda_0=1$, 
for any value of $s$, as required by trace preservation. 
For electrons ($s=\frac{1}{2}$), we  recover moreover the
eigenvalue $\lambda_1 = -\frac{1}{3}$ of the vector mode $K=1$
found heuristically in section \ref{heur.diag.sec}.

\subsection{Cooperon eigenvalues $\chi_K$}

From the vertex diagonalization \refeq{Xdiag}, we find the eigenvalues
\begin{eqnarray}
\chi_K &  =  \frac{X_{00}(K)}{\sqrt{2K+1}}
= \frac{1}{2K+1} \sum_q X_{Kq,Kq}  \\ 
&  = \frac{1}{2K+1} \sum_{q,\alpha\beta\gamma\delta}
 \cg{ss\alpha\beta}{Kq} X\abcd
\cg{ss\gamma\delta}{Kq}
\label{chiK.eq}
\end{eqnarray}
As a function of the elementary coupling coefficients $s_K$
defined in \refeq{def.sK}, the eigenvalues are 
\eqlab{ 
\chi_K =
(-)^{2s+K} \sum_{K'}  R_s(K',K) \, s_{K'}\ . 
} {G.xKsK}
Using \refeq{Rs}, the eigenvalues of the normalized spin-flip
crossed vertex \refeq{defX.spin-flip} with the only non-zero
coupling $s_1=d_s/3$ are thus 
\eqlab{
  \chi_K= (-)^{2s+K}(2s+1) \sixj{s}{s}{K}{s}{s}{1} = \frac{K(K+1)}{2s(s+1)}-1 \ .
}
{chiK_spin-flip_6j}
In the electronic case ($s=\frac{1}{2}$), this yields
indeed the previously found values $\chi_0= -1$ in the singlet
channel and $\chi_1 =\frac{1}{3}$ in the triplet channel.

\subsection{Direct recoupling of eigenvalues}
\label{rel.eigenvalues.sec}

The precise relation between diffuson and cooperon eigenvalues can be
understood by observing that both of them are obtained by a
recoupling procedure from a ``vertical'' coupling scheme of the
initial scattering amplitudes towards the relevant direction of
diagonalization, ``horizontal'' for the diffuson vertex, and
``diagonal'' for the cooperon vertex. 
This implies that the different eigenvalues are linked by simple recoupling
relations and useful sum rules.  

Starting from a scalar intensity vertex $V$ with no angular
dependence,  characterized by the product
$ \mv{V_{Kq} \overline{V_{K'q'}}} = \delta_{KK'}\delta_{qq'}\,s_K$,
the interaction vertices describing respectively
the ladder and the crossed diagrams can be written in the form
 \eqlab{
    \calL\abcd =   \sum_{K} s_K \, {\cal
T}_{\delta\beta,\gamma\alpha}^{(K)}, 
\qquad \qquad
    X_{\alpha\beta,\gamma\delta} =   \sum_{K} s_K \,  {\cal
    T}_{\beta\delta,\gamma\alpha}^{(K)}. 
}{verticalform} 
Clearly, the coupling constants  $s_K$ appear as the vertex
eigenvalues for the vertical coupling scheme. However, the projectors
are not in a form suitable for iteration of the multiple scattering
sequence with the horizontal product
\refeq{defproduit}. 
Using the
transformations 
\eqlab{
    \calT^{(K)}_{\delta\beta,\gamma\alpha} =  (-)^{K}  \sum_{K'}
    (-)^{2s+K'} \, R_s(K,K')\,
        \calT^{(K')}_{\alpha\beta,\gamma\delta}}{calTcalT'bis}
     \eqlab{  {\cal T}_{\beta\delta,\gamma\alpha}^{(K)}= \sum_{K'}
(-)^{2s+K'}\, R_s(K,K') \, T_{\alpha\beta,\gamma\delta}^{(K')}, 
}{calTT'} 
the vertices can be brought into the suitable form, 
 \eqlab{
    \calL\abcd =   \sum_{K} \lambda_K \, {\cal T}_{\alpha\beta,\gamma\delta}^{(K)}
\qquad \qquad
    X_{\alpha\beta,\gamma\delta} =   \sum_{K} \chi_K \,  {
    T}_{\alpha\beta,\gamma\delta}^{(K)}\ .
}{suitableform}
Using the definitions  \refeq{calTcalT'bis} and \refeq{calTT'}, 
the eigenvalues $\lambda_K$ and $\chi_K$ are then immediately derived as function of the vertical
eigenvalues $s_K$, given by expressions 
\refeq{lambdaK.sK} and \refeq{G.xKsK}. 
But this in turn implies that
the eigenvalues $\lambda_K$ and $\chi_K$ are also directly 
linked to each other by a simple
recoupling procedure.
Inverting the relations \refeq{lambdaK.sK} and \refeq{G.xKsK} 
with the help of the orthogonality relation
(\ref{orthoRs}) for $R_s$, one obtains the vertical eigenvalues as   
\begin{eqnarray}
s_K  & = & (-)^{2s+K} \sum_{K'} (-)^{K'}  R_s(K',K) 
\lambda_K  \label{sKof_lambda.eq}\\ 
 & = & (-)^{2s+K} \sum_{K'}  R_s(K',K) \chi_{K'}\ .
\label{sKof_chi.eq}
\end{eqnarray}
 Injecting  
\refeq{sKof_lambda} in  \refeq{lambdaK.sK} and \refeq{sKof_chi} into 
\refeq{G.xKsK}, we indeed find 
\eqlab{ 
\lambda_K = \sum_{K'} R_s(K',K)\,\chi_{K'}, \qquad  
\chi_{K} = \sum_{K'} R_s(K',K)\,\lambda_{K'} \ .
} {lambdaK.xK}
These relations had been derived previously in
the case of photon scattering (formula (52) of \cite{Mueller02},
$s=1$) and are here generalized to arbitrary spin. These
recoupling relations replace the heuristic prescription
$w_2\leftrightarrow w_3$ for the exchange of contraction weights
in the photonic case \cite{Mueller01} to arbitrary  spin.

Taking the trace \refeq{traceL} or \refeq{traceC} of the decompositions \refeq{suitableform} and
\refeq{verticalform}, we find the following useful sum rule:
\eqlab{ 
\fl 
\sum_K(2K+1)\chi_K =
\sum_K(2K+1)\lambda_K = \sum_K (2K+1)
\calT^{(K)}_{\beta\beta,\alpha\alpha}= (2s+1) s_0 \ .
}{supertrace} 
The last equality is explained by the fact that the ``horizontal'' trace over
all modes $K$ in \refeq{suitableform} projects onto the scalar
component $K=0$  in \refeq{verticalform}. By symmetry of our
recoupling relations, naturally also the inverse relation holds:
taking the ``vertical'' trace  in \refeq{verticalform} yields 
\eqlab{ 
\sum_K(2K+1)s_K = \calL_{\alpha\alpha,\beta\beta} = 
(2s+1) \lambda_0 \ .
}{veticaltrace} 
 
Let us demonstrate the power of these relations by taking again the spin-flip vertex
as a paradigmatic example. Its  
diffuson and cooperon eigenvalues  
\refeq{lambdaK_spin-flip_6j} and \refeq{chiK_spin-flip_6j} have turned
out to be  equal but of opposite sign,
$\chi_K(s) = - \lambda_K(s) $ for any $K$ and $s$. Indeed, 
comparing the eigenvalue expressions as a function of the elementary
couplings $s_K$, \refeq{lambdaK.sK} and \refeq{G.xKsK}, we can trace
back this sign to the fact that the spin-flip vertex has only one finite  
component, the vector coupling $s_1=d_s/3$. 
The relation \refeq{lambdaK.xK} then reduces
to an orthogonality of $6j$-symbols and yields immediately
$\lambda_K(s) = -\chi_K(s)$. The trace
\refeq{supertrace} reduces
to $\lambda_0+3\lambda_1=0$ (remember $s_0 = 0 $) which immediately fixes the triplet
eigenvalue to $\lambda_1=-\frac{1}{3}$ once the trace-preserving
eigenvalue $\lambda_0=1$ is known. By now it should be evident that these
values are after all of purely geometrical origin.

\subsection{Reciprocity}

Quite generally, the diffuson vertex of some microscopic spin
interaction $V$ is $\calL\abcd = \mv{V_{\gamma\alpha}
\overline{V_{\delta\beta}}} = \mv{V_{\gamma\alpha}
(V^\dagger)_{\beta\delta}}$ while the corresponding crossed vertex
is $X\abcd = \calL_{\alpha\delta,\gamma\beta} =
\mv{V_{\gamma\alpha} (V^\dagger)_{\delta\beta}}=
\mv{V_{\gamma\alpha} ([V^\mathrm{t}]^\dagger)_{\beta\delta}}$.
Clearly, these vertices are identical if the  microscopic
interaction is \textit{symmetric}, $V = V^\mathrm{t}$. This
complies with the general rule that the reciprocity
theorem assures perfect equality of ladder and crossed
contributions if the system's $S$-matrix is symmetric
\cite{BvT97}.
In terms of the irreducible amplitudes $V_{Kq}$ defined in
\refeq{VmicroKq}, symmetry of the microscopic interaction is
equivalently stated as
\eqlab{
V= V^\mathrm{t}  \quad
\Leftrightarrow \quad V_{Kq} = (-)^q V_{K\,{-q}}
}{cond-symm}
This
is to be contrasted with the hermiticity condition,
\eqlab{
V=V^\dagger \quad \Leftrightarrow \quad \overline{V_{Kq}} = (-)^q
V_{K\,{-q}}
}{cond_herm}
For example, a simple scalar interaction
$V^{(0)} = v_0 \mathbbm{1}$ is hermitian if $v_0$ is real.
Non-hermitian interaction would describe absorption or gain which
is known to preserve equality between ladder and crossed
contributions \cite{BvT97}.  Indeed, $V^{(0)}$ is always symmetric
since $V^{(0)} _{Kq} = \delta_{K0}\delta_{q0} \sqrt{2s+1} \, v_0$
fulfills \refeq{cond-symm}.

Less trivially, the spin-flip vector coupling $V^{(1)}  = g \,  \bi
J\cdot \bi S$ has irreducible components \eqlab{ V^{(1)} _{Kq} =
\delta_{K1} \frac{g}{\sqrt{c_s}} (-)^q J_{-q}. }{VKqreciproc} Of
course, the interaction is hermitian for a real $g$.  But
it is not symmetric in general because \refeq{VKqreciproc} is
different from $(-)^q V^{(1)} _{K\,{-q}}$ as soon as $J_q\ne (-)^q
J_{-q}$, i.e., if the impurities are not all aligned. This
is at the origin of the fact that spin-flip scattering suppresses
the diffusive pole of the cooperon  (the ``unitary case'' in
magnetoresistance \cite{Hikami80}). Furthermore, it explains the
observation that it is precisely the antisymmetric part
$t^{(1)}_{ij}$ of the atomic photon scattering tensor (i.e.,
the vector coupling component) that breaks  the equivalence of
ladder and crossed vertices \cite{Mueller01}. If, however, one can
align all vectorial scatterers, for example with an external
magnetic field, one can choose the quantization axis in this
direction such that $J_q = \delta_{q0} J_0$, and equality of
ladder and crossed terms is reestablished. This effect has been
observed with electronic transport in metal samples containing
magnetic impurities and subject to a strong magnetic field
\cite{Wash}. There, weak localization corrections to transport are
suppressed by spin-flip processes at low fields, but are restored
at high fields because all magnetic impurities are then aligned
with the field. A similar argument is at the origin of the
observed magnetic field enhancement of coherent backscattering of
light by a resonant sample of ultracold atoms \cite{Sigwarth04}
whose groundstate degeneracies are lifted by a magnetic field.

Having at hand the diffuson and cooperon eigenvalues
\refeq{lambdaK.sK} and \refeq{G.xKsK} as functions of the
elementary coupling coefficients $s_K$, we see immediately that a difference
between eigenvalues is generated only by coupling amplitudes $s_K$
with odd $K=1,3,\dots$ (for electrons and photons, only $K=1$ is
possible). A simple inspection of the microscopic interaction
vertex of a particular physical impurity type permits to decide
whether this coupling is of vectorial rank $K=1$, therefore breaks
the equivalence of ladder and crossed structures and eventually
leads to an effective dephasing of weak localization effects, or
whether it is of scalar or symmetric type $K=0,2$ and then does
not affect  localization effects.

\section{Consequences for relevant transport quantities}

\subsection{Diffuson spin transport}

The formalism of irreducible spin representations greatly simplifies
the expression of spin transport quantities that are of interest in
the growing number of ``spintronics'' applications. 
The probability of quantum diffusion, 
defined in subsection \ref{probadiff.sec}, with spin degrees of
freedom can be decomposed into its
irreducible components: 
\eqlab{
P \abcd 
(\bi q,\omega) 
= \frac{1}{2\pi\rho}
\mv{G^\mathrm{R}_{\alpha\gamma}
G^\mathrm{A}_{\beta\delta}}
= \sum_{K=0}^{2s} 
P_K(\bi q,\omega) \calT^{(K)}\abcd\ .
}{Pdecom}
The probability $P_K(\bi q,\omega)$ with spatial and temporal Fourier
variables $\bi q$ and $\omega$ can be 
computed independently in each spin sector $K$ if the
elementary scattering vertex as well as the average propagation
between scatterers has been diagonalised appropriately. 
For the summed ladder series $\mathcal{D}=\calL /(1-\sfG\calL)$, each
diffuson mode up to an overall normalisation reads 
\eqlab{
P^{(d)}_K (\bi q,\omega) 
= \frac{\lambda_K}{1-\lambda_K(1+\rmi \omega\tau -Dq^2 \tau)} 
}{Pqomega}
where the diffusion approximation for the intensity propagation in the 
Kubo limit $\omega\tau, q\ell\ll 1$ has been made; the diffusion constant in $d$ dimensions is 
$D=\ell^2/\tau d$ in terms of the scattering
mean-free path $\ell$ and the 
mean free time 
$\tau= \ell/v  = 1/(2\pi\rho \lambda_0)$ 
(evaluated from the self-energy in the Born approximation with $\lambda_0=
\sum_\gamma \calL_{\alpha\alpha,\gamma\gamma}=1$ in our
normalisation). 
The probability time dependence  for each irreducible sector 
therefore is of the form
$P_K (\bi q,t) \sim \exp[-Dq^2 t - t/\tau_d(K)]$ with a diffuson spin
decay time 
\eqlab{
\tau_d(K) = \tau \frac{\lambda_K}{1-\lambda_K}
}{taudK}
given directly as function of the vertex eigenvalue $\lambda_K$. 
The time-dependent diffuson probability of classical scattering
behaves as 
\eqlab{
P^{(d)}\abcd 
(\bi q,t)
= \sum_{K=0}^{2s} \rme^{-Dq^2 t - t/\tau_d(K)} \calT^{(K)}\abcd\ .
}{P_time}

After injection of a certain spin state $\ket{s\alpha}$ at $\bmr$, 
the total classical probability for final states with arbitrary spin
$\gamma$ at arbitrary position $\bmr'$ should be normalised to unity, $
 \sum_\gamma P_{\alpha\alpha,\gamma\gamma}(\bi q=0,t) = 1$. Indeed, 
the trace over the final spin index $\gamma$  in \refeq{P_time} projects onto the  scalar
component $K=0$, such that the conservation of probability requires
$1/\tau_d(0)=0$ which is indeed the case for a unit scalar eigenvalue
$\lambda_0=1$. 

The overall probability of retaining the initial spin state, say
$\alpha=+\frac{1}{2}=:+$, 
is $P_{++,++}(t) = (\calT^{(0)}_{++,++}) +
\rme^{-t/\tau_1}  (\calT^{(1)}_{++,++}) =
\frac{1}{2}(1+\rme^{-t/\tau_d(1)})$ 
which deviates only for short times from the equidistribution value
$\frac{1}{2}$. The degree of spin polarisation $\pi(t)= P_{++,++}(t)-P_{++,--}(t)$
relaxes on the time scale $\tau_d(1)$ as  
$\pi(t)  =  \rme^{-t/\tau_d(1)}$. 

Consider for the
sake of concreteness the case of multiple electronic spin-flip
scattering with the vertex \refeq{defV0.spin-flip}. This vertex has a negative
eigenvalue $\lambda_1=-\frac{1}{3}$ such that formally the decay time
$\tau_d(1)$ becomes negative, which renders the above predictions 
inacceptable. In fact, in a disordered electronic sample, there are
scalar defects responsible for elastic scattering and momentum relaxation,
say with a rate $\gamma_e=n_ev_0^2$ depending on the density $n_e$ of
defects and their 
interaction strength $v_0$. In spin space, these vertices have unit
eigenvalues both for ladder and crossed vertices. In addition there are magnetic
impurities, each with a scattering vertex
\refeq{defV0.spin-flip}, centered at random positions $\bmr_m$ with
density $n_m$ and
corresponding momentum scattering rate $\gamma_m=n_mg^2 \frac{1}{3}J(J+1)s(s+1)$. The inverse of the total
scattering rate $\gamma=\gamma_e+\gamma_m$ is the 
mean-free time $\tau=1/\gamma$. The effective average spin-flip
scattering vertex then has the normalised ladder eigenvalues 
\eqlab{
\lambda_K^{\mathrm{eff}} =\frac{ \gamma_e + \gamma_m
\lambda_K}{\gamma}
=1- \frac{\gamma_m}{\gamma}(1-\lambda_K) 
}{lambdaeff}
which for the spin-flip case are $\lambda_0^{\mathrm{eff}} =1$ and
$\lambda_1^{\mathrm{eff}} =1- 4\gamma_m/3\gamma$. The characteristic spin polarisation
decay rate reads 
\eq{
\frac{1}{\tau_d(1)} =
\frac{1-\lambda^{\mathrm{eff}}_1}{\tau\lambda^{\mathrm{eff}}_1} = 
\frac{ 1-\lambda_1}{\tau_m} + O(\frac{\gamma_m}{\gamma}) \approx
\frac{4}{3\tau_m} \ .
}

\subsection{Cooperon dephasing}

The cooperon is the sum of all maximally crossed diagrams which,
strictly speaking, starts with the second-order scattering term
because the single scattering event is already counted in the
diffuson: $C=X \sfG X/(1-\sfG X)$. Here, $X$ is the crossed
vertex associated to the diffuson vertex for scattering with a total
rate $\gamma=1/\tau$ and normalised to a unit diffuson eigenvalue $\lambda_0=1$. 
Summing the geometric series with
the returned advanced propagator line (see figure \ref{cooperon.fig}(c))
gives the decomposition 
\eqlab{
C\abcd(\bQ,\omega)= \sum_{K=0}^{2s}  C_K(\bQ,\omega) T^{(K)}\abcd
}{cooperon_decomp}
where $\bQ = \bi k+\bi k'$ is the sum of external momenta. 
The cooperon eigenfunctions for each irreducible mode are 
\eq{
C_K(\bQ,\omega) = \frac{1}{\tau}\frac
{\chi_K}{-\rmi\omega + DQ^2  +1/\tau_c(K)}
}
with characteristic cooperon dephasing times
\eq{
\tau_c(K) = \tau \frac{\chi_K}{1-\chi_K}
}
that depend directly on the crossed eigenvalues $\chi_K$.  

In applications to weak localization, one needs the integrated
cooperon, both over momenta and spin indices, that counts arbitrary
loops of counter-propagating amplitudes. Up to a normalisation, 
\eqlab{
P_c(\omega)= \sum_\bQ \sum_\gamma
C_{\alpha\gamma,\gamma\alpha}(\bQ,\omega) = \sum_{K=0}^{2s}  w_K \sum_\bQ C_K(\bQ,\omega)
}{intcooperon}
where the cooperon spin-channel weights $w_K$ are determined by the
crossed contraction 
$w_K = \sum_\gamma  T^{(K)}_{\alpha\gamma,\gamma\alpha}$ that
corresponds to the sum over all final spin
states $\beta=\gamma$ in the maximally crossed diagrams of figure
1(b). 
Using the recoupling formula \refeq{transfbase}, one finds 
\eq{
w_K = \sum_{K'}R_s(K,K') \sum_\gamma
\calT^{(K')}_{\alpha\alpha,\gamma\gamma} = R_s(K,0) = (-)^{2s+K}
\frac{2K+1}{2s+1}
\ .
}
Remarkably, although the integrated cooperon describes a
renormalisation of intensity diffusion (the scalar diffuson mode
$K=0$), in general all the cooperon modes $K=0,\dots,2s$ contribute with non-zero
weights 
\cite{kushnir_masterthesis05}. 

Taking again the case of electronic elastic and spin-flip scattering
as an example, the singlet and triplet channel weights are $w_0=-\frac{1}{2}$ and
$w_1=\frac{3}{2}$, and the effective eigenvalues are 
\eqlab{
\chi_K^{\mathrm{eff}} =\frac{ \gamma_e + \gamma_m
\chi_K}{\gamma}
=1- \frac{\gamma_m}{\gamma}(1-\chi_K) \ .
}{chieff}
For a small spin-flip scattering rate $\gamma_m/\gamma\ll 1$, the
corresponding dephasing rates are 
\eq{
\frac{1}{\tau_c(K)} =\frac{1-\chi_K }{\tau_m} 
\ .}
Inserting the spin-flip eigenvalues $\chi_0=-1$ and $\chi_1=\frac{1}{3}$, 
the singlet and triplet dephasing times
 are $\tau_c(0)= \tau_m/2$ and $\tau_c(1) = 3\tau_m/2$. 
If the 
eigenvalues \refeq{chieff} in the numerator of \refeq{intcooperon} are
approximated by unity, the integrated spin-flip
cooperon finally reads 
\eqlab{
P_c(\omega)= \sum_\bQ \left[\frac{3}{2}\,
\frac{1}{-\rmi\omega + DQ^2  + 2/3\tau_m}  - \frac{1}{2}\,
\frac{1}{-\rmi\omega + DQ^2  + 2/\tau_m}
\right].
}{spinflip_cooperon}
The coupling to the uncontrolled degrees of freedom of magnetic
impurites dephases both the
singlet and triplet channels irreversibly and leads to a drastic decrease of weak
localization effects in disordered electronic samples. 
A similar effect is found for the weak localization of photons 
scattered by cold atoms with degenerate
dipole transitions \cite{amm}.

Note that in the case of electronic spin-orbit scattering, the triplet channel
with positive weight $w_1=\frac{3}{2}$ is also rapidly damped,
whereas the singlet channel with negative weight
$w_0=(-)^{2s}/(2s+1) = -\frac{1}{2}$ survives and leads to antilocalization and a
positive magnetoresistance \cite{bergmann84} which appears here as characteristic
for any half-integer spin.

\section{Summary and conclusion}

In this paper, we have developed a systematic method to diagonalize
the elementary spin scattering vertices which are the building block of
diffuson and cooperon multiple scattering sequences for particles of
arbitrary spin $s$. Our results therefore provide the conceptual background for a
truly unified description of the mesoscopic spin physics of electrons and
photons.  
 We have identified the relevant projectors 
onto invariant subspaces that are irreducible with respect to the rotation
group. Once these operators have been obtained, the
diagonalization allows us to transform the vertical coupling
scheme for the scattering amplitudes into a horizontal
scheme necessary for subsequent
iteration of the multiple scattering sequence. We have obtained the diffuson
and cooperon scattering eigenvalues 
as a function of the microscopic
scattering mechanism, together with 
simple recoupling relations as well as useful sum
rules. 
We have shown how these eigenvalues directly enter the expressions of the phase
coherence times of weak localization. 

 The method presented here may be
extended to non-scalar vertices such as the transverse photon
propagator $\mathsf{G}(\bi q)$ that was diagonalized exactly in
\cite{Mueller02}. This transverse propagator is no longer purely
scalar at non-zero momentum transfer $\bi q \ne 0$, but contains
quadrupolar parts coupling the modes $K=0,2$. 
A treatment of these non-scalar vertices would start from the
expressions \refeq{VLmdef} and \refeq{XLmdef} of the present work and derive the
appropriate projectors and eigenvalues. 
On a similar line of
thought, light scattering by nematic crystals \cite{Sapienza04} as
well as photon scattering vertices of atoms under the influence of
an external magnetic field \cite{Sigwarth04}  require anisotropic
diagonalization.  Finally, these techniques may become useful for
quantum computation schemes involving spin degrees of freedom
(such as the one studied by Loss and di Vincenzo \cite{Loss98}, to
cite a paper employing a superoperator formalism quite similar to
ours) or for entanglement characterization in irreducible
representations of observable-induced tensor product spaces
\cite{Zanardi04}.

\ack
CAM acknowledges a short term research grant by the Minerva Foundation
and the hospitality of EA at the Technion where the manuscript  was brought into
final form. The authors acknowledge financial support by the
\textsc{Procope} program of MAE and DAAD, the Israel Academy of
Science, the Fund for Promotion of Reserach at the Technion, 
as well as the GDR 2426 Physique
Quantique M\'esoscopique.  

\appendix
\section{Photon scattering}
\label{photon.sec}

\subsection{Atomic vertex eigenvalues}

For photons, polarisation-dependent scattering proves
more complicated than for electrons because of field transversality. The spin degrees
of freedom are \textit{not} decoupled from average propagation such that
the complete diagonalization of the diffuson and cooperon series is
much more involved.   
Nonetheless, the projection onto irreducible subspaces 
permits to derive all eigenvalues and projectors for isotropic photon
vertices as well, as has been donc in \cite{Mueller02} for the case of
photon scattering from degenerate atomic dipole transitions. 
For resonant photon scattering, the elementary interaction is of the form
$V_\mathrm{dip}=-\bi D\cdot \bi E$ where the electric field operator $\bi E$,
proportional to the polarisation vector $\bvarepsilon$, creates or
annihilates photons whereas the dipole operator $\bi D$ induces
internal transitions between electronic states with angular momentum
$\Jg$ and $\Je$. Obviously,
the elementary interaction $V_\mathrm{dip}$ is of the same vectorial type as the
electronic spin-flip vertex \refeq{Vm}. The full photon scattering
process, however, comprises the annihilation of the incident photon
followed by the creation of the scattered one. Therefore, the full
scattering amplitude is a second-order process in $V_\mathrm{dip}$, and
its elementary coupling coefficients $s_K$ (see eq.~(30) of
\cite{Mueller02}) contain already all orders
$K=0,1,2$ obtained by the recoupling of two vector interactions of
rank one. The ladder and crossed eigenvalues are then expressed in terms of
$6j$- and $9j$-symbols. All relations between eigenvalues derived in section
\ref{eigenvalues.sec} apply to the photon case as well. The sum rule
\refeq{supertrace} has not been evaluated so far. For resonant photon
scattering from a closed dipole transition $\Jg \to
\Je$, it reads  
\eq{
\sum_K(2K+1)\chi_K =
\sum_K(2K+1)\lambda_K = \frac{3(2\Je+1)}{2\Jg+1} \ .
}

The more general case of photon scattering from entire multiplets of
hyperfine or fine structure dipole transitions can be treated as
well \cite{hyperfine}. In that case, the eigenvalues become frequency dependent, but all
algebraic relations of the present work continue to hold. 

\subsection{Photon projectors} 
\label{gram-schmidt.sec}

Concerning the projectors onto irreducible subspaces, it  
was shown in \cite{Mueller02} that one and the same set of projectors   
\begin{eqnarray}
   \calT^{(0)}_{il,jk} & 
= & \frac{1}{3}\,\delta_{il}\delta_{jk},
\label{T0.photon.eq}\\
   \calT^{(1)}_ {il,jk}&  
= &\frac{1}{2}\,(\delta_{ij}\delta_{kl} - \delta_{ik}\delta_{jl}),  \label{T1.photon.eq}\\
  \calT^{(2)}_{il,jk} &  
= &\frac{1}{2}\,(\delta_{ij}\delta_{kl} + \delta_{ik}\delta_{jl})
-
  \frac{1}{3}\,\delta_{il}\delta_{jk}.
\label{T2.photon.eq}
\end{eqnarray}
diagonalises both ladder and crossed vertex alike. 
Here, the indices $i,j,k,l$ are Cartesian indices in configuration
space $\mathbbm{R}^3$. 

We still need to connect these tensors to the diffuson
(super-)projectors 
$\calT^{(K)}$ derived  in  \ref{diffusonprojectors.sec}
and the seemingly different cooperon projectors $T^{(K)}$ 
derived in  \ref{cooperonprojectors.sec}.  
Of course, all formulae of the present work are
designed to apply to spin 1 as well. But spin 1 is special because its states
transform under the fundamental representation SO(3) itself.
This representation has the peculiar feature that its dimension is
$3$ and therefore equal to the dimension of the abstract group.
For spin 1, one therefore can use
either the standard spherical basis  $\{\ket{1q}\}$ where
$S^z\ket{1q} = q\ket{1q}$ is diagonal (and which was used
throughout the present paper) or the Cartesian basis of
$\mathbbm{R}^3$ itself where the generators are $S^j_{kl} = -i
\epsilon_{ijk}$ (in this so-called adjoint representation, the
generators are essentially given by the structure constants
$f_{jkl} = -\epsilon_{jkl}$ of the Lie algebra so(3)). Using the
latter representation in the definitions \refeq{calT0} and \refeq{calT1}, we
indeed recover immediately \refeq{T0.photon} and
\refeq{T1.photon}.


In order to obtain the symmetric traceless projector $K=2$, we have to
push the calculation one step further. Extrapolating from the cases
$K=0,1$, the construction rule for higher-rank projectors
\refeq{calTK} should be clear: they are complete contractions 
\eqlab{ 
\calT^{(K)}
= \sum_{i_1i_2\dots i_K} O^{i_1i_2\dots i_K}
\psopdot{O^{i_1i_2\dots i_K}} 
}{calTKdeX} 
of operators 
$O^{i_1i_2\dots i_K}$ that are direct products of $K$ copies of
$S^i$ with the appropriate symmetrisation. Up to
normalisation, the operator of the $K=2$ projector is 
\eq{ 
O^{ij} = S^iS^j - \sum_k \psop{O^k}{S^i S^j}O^k -
     \psop{O^0}{S^i S^j} O^0
} 
where $\sqrt{d_s} \, O^0 = \mathbbm{1}$ pertains to the scalar
projector \refeq{calT0}. This construction is nothing but the
Gram-Schmidt procedure used to orthogonalize the basis $\{e_n\}$
of a vector space according to $e_1\mapsto e_1$, $e_2\mapsto e_2-
(e_1\cdot e_2)e_1$, \textit{etc}. After short algebra, one finds
\eq{ O^{ij}= \frac{1}{2}(S^iS^j+S^jS^i) -
\frac{s(s+1)}{3}\,\delta_{ij}\,\mathbbm{1} } without need for
further normalisation. This operator is manifestly symmetric and
traceless by construction. Using $S^j_{kl} = -i
\epsilon_{ijk}$, we obtain the traceless symmetric projector \refeq{T2.photon}
we sought to justify.  

For particles of larger spin than $s=1$ (or more general vertices with
higher-order irreducible components), the calculation of higher-order
projectors requires to compute 
traces of increasingly large products of spin operators, $\tr\{S^i
S^j \dots S^p\}$. Beyond the first terms $K=0,1,2$, carrying out
the traces becomes rapidly cumbersome, and the formulation
\refeq{TKqabcd} in terms of spherical components $T^{(K)}_q$ turns
out to be more economic since the CG-coefficients automatically
incorporate the correct symmetrisation.


At this point, we have completely justified the diffuson tensors
\refeq{T0.photon}--\refeq{T2.photon}.  
However, at first sight the corresponding crossed projectors
$T^{(K)}$ 
derived in  \ref{cooperonprojectors.sec}
have nothing in common with them.
But inserting the relevant transformation
coefficients  for $s=1$, 
\eq{
(R_1)_{K,K'} = 
\left(
    \begin{array}{ccc}
            1/3  & -1/3  & 1/3 \\
            -1 & 1/2 &  1/2 \\
            5/3  &  5/6 & 1/6
    \end{array}
\right)
}
into the recoupling relation \refeq{TKdecalTK'} yields indeed 
$T^{(K)}_{il,jk}  = \calT^{(K)}_{il,jk}$, $K=0,1,2$.
This proves
that the diagonalization of spin 1 ladder and crossed vertices
in Cartesian components involves the unique set of isotropic
projectors \refeq{T0.photon}-\refeq{T2.photon}. 
This is no longer the case for half-integer spins
$s=\frac{1}{2},\frac{3}{2},\dots$ because their SU(2)
representations are complex unitary which means that ladder and
crossed vertices couple rotationally different objects. The same
conclusion holds for integer spins $s=2,4,\dots$, though they
admit real orthogonal representations of SO(3). This is because
the adjoint representation is not available and one has to work
\textit{a priori} with two  distinct sets of projectors.

\end{document}